\documentclass{jfm}

\usepackage{epstopdf, epsfig}
\usepackage{natbib}
\usepackage{amsmath}
\usepackage{amssymb}

\usepackage{graphicx}
\usepackage{subcaption}
\usepackage{dcolumn}
\usepackage{bm}
\usepackage{marvosym}
\usepackage{epstopdf}

\usepackage{color}
\definecolor{light-gray}{gray}{0.5}
\definecolor{blue}{rgb}{0.0,0.0,1.0}
\definecolor{green}{rgb}{0.0,0.5,0.0}
\definecolor{red}{rgb}{1.0,0.0,0.0}
\definecolor{cyan}{rgb}{0.0,0.75,0.75}
\definecolor{magenta}{rgb}{0.75,0.0,0.75}
\definecolor{yellow}{rgb}{0.75,0.75,0.0}

\newcommand{\avg}[1]{\langle{#1}\rangle}

\newcommand{\grad}{\bm \nabla}

\newcommand{\lt}{\left}
\newcommand{\rt}{\right}

\newcommand{\red}[1]{\textcolor{red}{#1}}

\newcommand{\rmc}{Rm_c^{\mathrm{turb}}}

\shorttitle{Rotationally induced coherence in turbulent kinematic dynamos}
\shortauthor{V. Dallas and S. M. Tobias}

\title{Rotationally induced coherence in \\ turbulent kinematic dynamos}

\author{Vassilios Dallas\aff{1}\corresp{\email{vassilios.dallas@maths.ox.ac.uk}} and Steven M. Tobias\aff{2}}

\affiliation{\aff{1} Mathematical Institute, University of Oxford, Woodstock Road, Oxford OX2 6GG, UK
\aff{2} Department of Applied Mathematics, University of Leeds, Leeds LS2 9JT, UK}

\begin{document}

\maketitle

\begin{abstract}
We consider rotating, kinematic dynamos at low magnetic Prandtl number $Pm$. We show that the inclusion of rotation leads to an increase in spatio-temporal coherence and a modification of the turbulent spectrum. These effects make the flow more efficient in driving the dynamo, in the sense that the energy injection rate required to reach the critical value of the magnetic Reynolds number $Rm_c$ is reduced in comparison with a non-rotating dynamo \citep{sda17}. For random dynamos it is known that the growth-rate would largely be determined by the spectral index of the flow at the resistive scale. Here, however, we demonstrate that the dynamo growth-rate in rotating flows is increased by the rotationally induced long-lived large scale eddies with a coherence time greater than the local turnover time. These eddies play the major role in determining the dynamo growth-rate.
\end{abstract}

\maketitle

\section{Introduction}

The origin of magnetic fields in planets, stars and galaxies is often attributed to hydromagnetic dynamo action \citep{parker79}. In a dynamo,  magnetic field is generated against the action of Ohmic dissipation by stretching within the flow. In many cases of astrophysical and geophysical interest, and indeed for liquid metal experiments, the dynamo is generated by a fluid or plasma where the magnetic Prandtl number $Pm = \nu/\eta$, with $\nu$ the viscosity and $\eta$ the diffusivity of the fluids,  is small and so the Reynolds number is much larger than the magnetic Reynolds number. This has important consequences both for the nature of the magneto-turbulence should the magnetic field be successfully generated \cite[see e.g.][]{favier2012} and perhaps more fundamentally for the efficiency of dynamo action itself \citep[as discussed in][]{moffatt1970,scheketal2007,tobias2011mhd}.

In this study, we address a fundamental question of  kinematic dynamo action in a low $Pm$ fluid, which is concerned with what determines the growth rate of the dynamo instability in turbulent flows at the early stages of the dynamo. Turbulence is difficult in that it requires a description of the coexistence of flows that remain coherent on long timescales (so-called coherent structures often taking the form of vortices) and a component that has short correlation times and can be considered `random'.  As we shall discuss the random component may be described by a statistical theory predicated on the random superposition of flows. The coherent structures however require a different theory for the generation of magnetic fields. Of course in a turbulent flow that has both components i.e. a flow that has a well-defined spectrum with a random component  and a more coherent component associated with some constraint, such as rotation, stratification or  shear, the competition between the two components in generating magnetic field is of significant interest. Such flows are ubiquitous in geophysics and astrophysics.

The evolution of the magnetic field $\bf B$ is given by the linear induction equation (see Eq. \eqref{eqn:induction} below), which, for steady and oscillatory flows admits solutions of the form ${\bf B} = {\bf b}({\bf x})\exp(\lambda t)$, where $\lambda$ is a complex eigenvalue and $\gamma = Re(\lambda)$ is the growth rate of the magnetic field. For turbulent statistically steady flows $\gamma$ represents the mean growth-rate of the dynamo. For random kinematic dynamos i.e. those with short correlation times, it is generally believed that there is a straight-forward relationship between the statistical properties of turbulence and the dynamo growth rate $\gamma$. In particular, for the Kazantzev-Kraichnan model \citep{Kazantsev68,Kraichnan68}, which considers the kinematic dynamo instability driven by a random velocity field that is homogeneous, Gaussian and $\delta$-correlated in time, $\gamma$ is determined by the exponent of the energy spectrum in the neighbourhood of the dissipative scale \citep[see e.g.][for a review of such dynamos]{tobias2011mhd}. The asymptotic analysis of the Kazantzev-Kraichnan model shows that the growth time $\tau_\gamma$ is of the order of the turnover of the eddies $\tau_{_{NL}}$ at the resistive scale $\ell_\eta$ \citep{boldyrevcattaneo04}. This makes sense as these eddies have the highest shear rates in such flows. 
For random dynamos at low $Pm=Rm/Re$, such as those in liquid metals or stellar interiors, where the kinetic Reynolds number $Re = UL/\nu$ is much bigger than the magnetic Reynolds number $Rm = UL/\eta$ (with $U$ the rms velocity, $L$ the size of the computational domain, $\nu$ the kinematic viscosity and $\eta$ the magnetic diffusivity), the growth-rate is then completely determined by the spectral slope of the velocity at the dissipative scale of the magnetic field. 
However, for geophysical and astrophysical flows, which have a substantial coherent component, outside of the range of validity of the Kazantsev-Kraichnan model it may be that characteristics other than the spectral slope of the velocity field do play a key role in determining the threshold of the dynamo instability.

\begin{figure}
 \centering
 \begin{subfigure}{0.45\textwidth}
   \includegraphics[width=\textwidth]{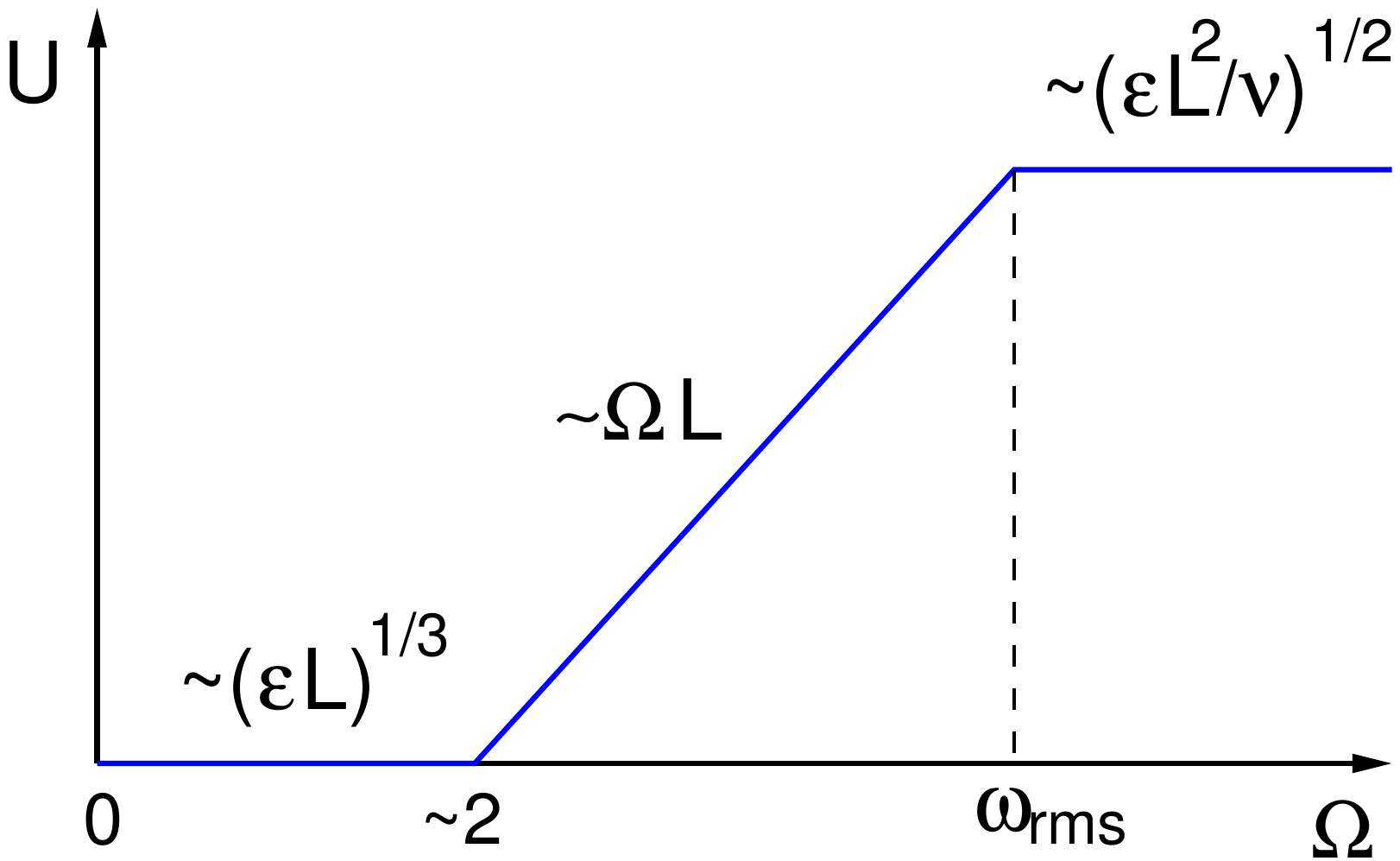}
   \caption{}
   \label{fig:energy}
 \end{subfigure} 
 \quad
 \begin{subfigure}{0.45\textwidth}
   \includegraphics[width=\textwidth]{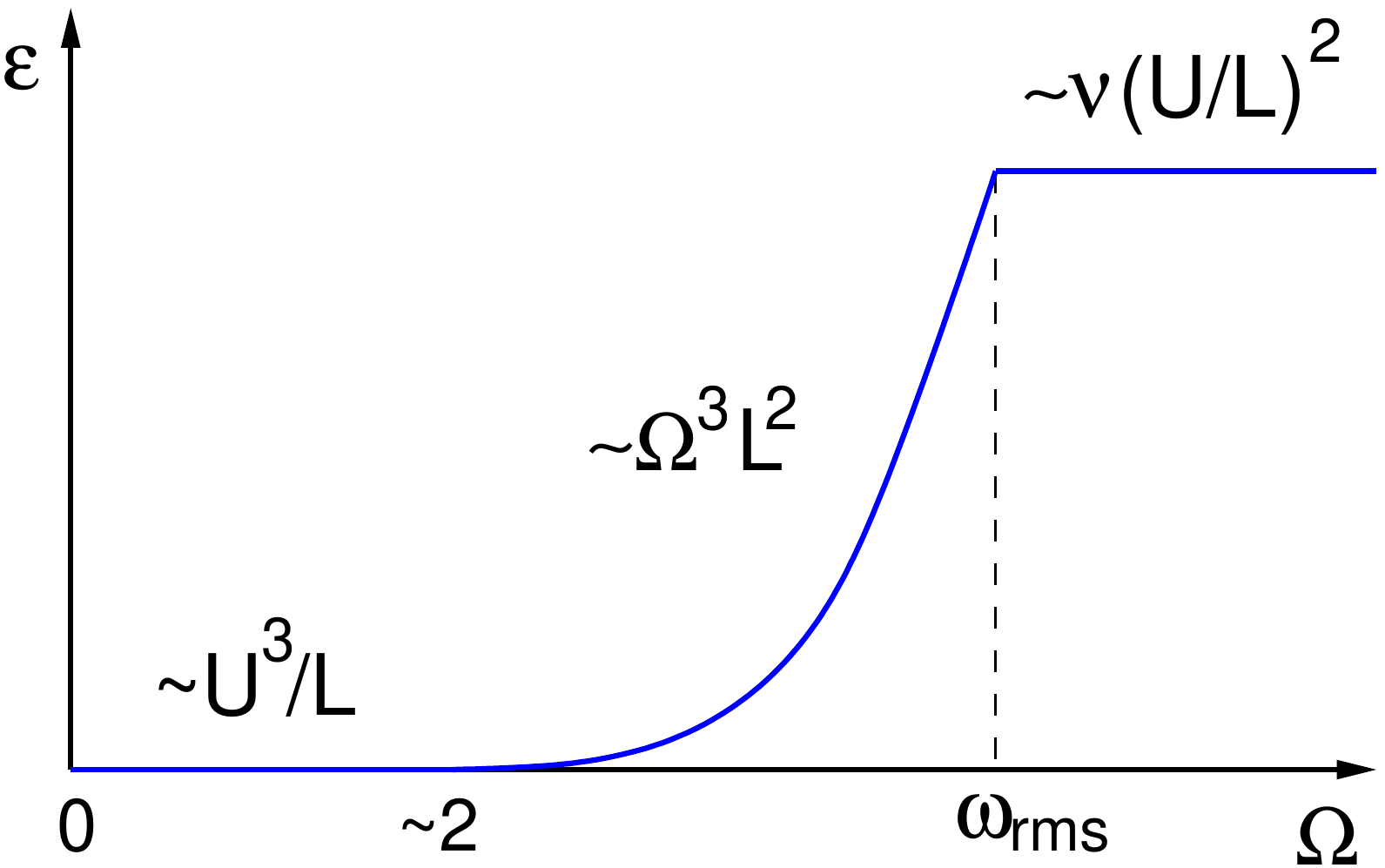}
   \caption{}
   \label{fig:dissipation}
 \end{subfigure}
 \caption{(Color online) The scaling regimes for a) the rms velocity $U$ and b) the dissipation rate $\epsilon$ of kinetic energy in terms of the rotation rate $\Omega$} 
 \label{fig:scalings} 
 \end{figure}

Here we consider turbulent flows under the effect of background rotation. Recently, \cite{sda17} demonstrated that the dynamo threshold can be significantly reduced if the flow is submitted to background rotation. The flows that were considered to show this, fall in three scaling regimes, which are illustrated in Fig. \ref{fig:scalings} and summarised below. In regime I ($0 < \Omega \leq 2$), for sufficiently small rotation the flows are random and the underlying flow is not far away from 3D isotropic turbulence. Here the rms velocity $U \propto (\epsilon L)^{1/3}$ and the dissipation rate $\epsilon \propto U^3/L$ for $Re \gg 1$. 
In regime II ($2 < \Omega < \omega_{rms}$), for moderate rotation the flows consist of two components; a coherent and a random component. The flows here are anisotropic, with fluctuations being suppressed along the direction of rotation. In this regime, there is an inverse cascade  that forms large scale coherent vortices, called condensates. 
Similar large-scale vortices have been identified in rapidly-rotating convection, first in reduced models \citep{Juletal2012,Rubetal2014} and then in DNS \citep{Stelletal2014, ghj2014, fsm2014}. The condensates are thus believed to be robust features of rotating turbulence.
The growth of the condensate saturates when the counter-rotating vortex locally cancels the effect of global rotation with $U \propto \Omega L$ \citep{bartello1994coherent,alexakis15}. In this case, the scaling of the dissipation rate is $\epsilon \propto \Omega^3 L^2$, since $\epsilon \propto U^3/L$ and $U \propto \Omega L$ reaching a lower finite value (in comparison to regime I), independent of $Re$ at $Re \gg 1$. 
Finally, in regime III ($\Omega \geq \omega_{rms}$) for large rotation the inverse cascade saturates owing to viscous forces and 
dissipative effects are dominant. Thus, $U \propto (\epsilon L^2/\nu)^{1/2}$ and $\epsilon \propto \nu (U/L)^2$, similar to the energy condensation at large scales of 2D turbulence \citep{boffettaecke12}.

It was found that increasing the effects of rotation in the flow leads initially to a hindering and then a facilitation of the dynamo properties of the flow at low $Pm$; in particular it was argued that the presence of rotation makes such a low $Pm$ dynamo act more like a high $Pm$ dynamo. In this paper we investigate the reason for this transition in detail; we determine that this occurs owing to an increase in the coherence of the dynamo eddies, rather than a modification of the spectral slope (which would be important if the coherence of the eddies remained small).
 
%

\section{Numerical set-up}

The dimensional governing equations for the kinematic dynamo problem in a rotating frame of reference are
\begin{align}
\partial_t {\bf u} + ({\bf u} \cdot \bm{\nabla}) {\bf u} + 2 \bm{\Omega} \times {\bf u} = & - \frac{1}{\rho}\bm{\nabla} P + \nu \bm \grad^2 {\bf u} + \bf f, 
\label{eqn:NS}
 \\
\partial_t {\bf B} = & \,\bm{\nabla} \times \left( {\bf u} \times {\bf B} \right) + \eta \bm \grad^2 {\bf B},
\label{eqn:induction}
\end{align}
where $\bf{u, B}$ are the velocity and the magnetic field respectively with $\bm{\nabla} \cdot {\bf u} =\bm{\nabla} \cdot {\bf B} = 0$, \newline
$\nu$ is the kinematic viscosity,
$\eta$ is the magnetic diffusivity,
$P$ is the reduced pressure due to the centrifugal acceleration, and 
$\rho$ is the mass density.
The background rotation is given by $\bm{\Omega} = \Omega \hat{\bf e}_z$.
We integrate these equations numerically in a Cartesian cubic periodic box of length $2 \pi L$
using a pseudo-spectral code 
with a third-order Runge-Kutta scheme for the time advancement and the 2/3 dealiasing rule; 
for more details of the numerical code see \cite{mpicode05b}. 
The force density is chosen to be a 2.5D {\it non-helical} Roberts flow 
${\bf f} = f_0 \left( \cos(k_f y), \sin(k_f x), \cos(k_f y) + \sin (k_f x) \right)$,  
where the forcing wavenumber $k_f L = 4$ \citep{roberts1972}; this is a cellular forcing with no net helicity.

In this study, we are interested in the low $Pm$ limit. To model the $Pm = Rm/Re \ll 1$ (or the $Re \gg 1$ limit) we use hyperviscosity $\nu_h$ where the Laplacian in the Navier-Stokes equation Eq. \eqref{eqn:NS} is changed to $\grad^8$. 
The use of hyperviscosity assumes that the large scales of the flow do not depend on the exact mechanism that energy is dissipated in the small scales,
and thus in principle should always be compared to the results of large $Re$ simulations. Regular Ohmic dissipation ($\eta \nabla^2 {\bf B}$) is used in the induction equation, as it must for dynamo calculations.

For such a forced system, many of the familiar non-dimensional numbers may only be determined \textit{a posteriori}. For example, non-dimensional parameters based on the energy injection rate $\epsilon = \avg{{\bf u} \cdot {\bf f}}$ (where $\left\langle \cdot \right\rangle$ denotes volume and time average) 
are defined as follows.
The magnetic Reynolds number is $Rm = (\epsilon/k_f)^{1/3}/(k_f \eta)$, 
the Rossby number is $Ro = (\epsilon/k_f)^{1/3} k_f /(2 \Omega)$ and due to the use of hyperviscosity the kinetic Reynolds number is $Re = (\epsilon/k_f)^{1/3}/(k_f^7 \nu_h)$.
%
The values of these parameters are given in Table \ref{tbl:parameters}.
For comparison we also provide the values of the non-dimensional parameters based on the rms velocity $U = \langle |{\bf u}^2| \rangle^{\frac{1}{2}} $ of the flow, $Re_{_U}=U/(k_f^7 \nu_h)$ and $Ro_{_U}=U k_f/(2 \Omega)$ (see Table \ref{tbl:parameters}).

 \begin{table}
  \centering
    \begin{tabular}{c|ccccccc} 
   $\Omega$  &   $Ro$     &    $Re (\times 10^{9})$  &  $Ro_{_U}$ &   $Re_{_U}(\times 10^{10})$ & N     & $\rmc$ \\ \hline
      0      &   $\infty$ &    $64.6$                  &  $\infty$  &   $18.0$                     & 512   & 23.6   \\ 
      1      &   $1.0$    &    $57.1$                  &  $3.0$     &   $18.5$                     & 512   & 34.9   \\  
      3      &   $0.2$    &    $41.4$                  &  $3.5$     &   $64.5$                    & 512   & 1.81   \\
    \end{tabular}
  \caption{Numerical parameters of the simulations. For all runs $f_0=1$, $L=1$ and $k_fL=4$. The reported values are based on hyperviscosity. The value of the turbulent critical magnetic Reynolds number is defined as $\rmc \equiv \lim_{Re\rightarrow\infty}Rm_c$.}
  \label{tbl:parameters}
\end{table}

 \section{Multiscale dynamics}

Much of what is known today in dynamo theory is related to single-scale dynamos, which are characterised by a single magnetic Reynolds number $Rm = U \ell /\eta$, where $U$ is the characteristic flow velocity and $\ell$ is the
single length scale of the flow. For these flows we know that the dynamo instability occurs for a critical value of the magnetic Reynolds number $Rm_c$. Then, as $Rm$ increases we can observe two distinct types of dynamos; fast dynamos 
with $\lim_{\eta \to 0} \gamma(\eta) = \gamma_0 > 0$ and slow dynamos with $\lim_{\eta \to 0} \gamma(\eta) = \gamma_0 \leq 0$.
 Note that, for flows defined at a single scale $\ell$, the natural unit of 
$\gamma$ is the inverse of the eddy turnover time $\tau_{_{NL}} = \ell/U$.

One way to extend some of these ideas to turbulent flows characterised by multiple scales is to define the scale-dependent magnetic Reynolds numbers $Rm(k) = u(k)/(\eta_c k)$ \citep[see e.g.][]{tobiascattaneo08a}. Here $\eta_c$ is the value of the magnetic diffusivity at the dynamo onset. One may also define the turnover time $\tau_{_{NL}}(k) = 1/(k u(k))$ and consider a multiscale velocity field $u(k)$ which can be determined from 
\begin{equation}
 u(k) = (kE_u(k))^{1/2}
 \label{eq:velk}
\end{equation}
using the definition of the kinetic energy spectrum. 
Thus, by computing $E_u(k)$ we can easily obtain the scale-dependent quantities $Rm(k)$ and $\tau_{_{NL}}(k)$, which is a good starting point to understand the factors that determine the dynamo growth rate in turbulent flows.

\begin{figure}
 \begin{subfigure}{0.5\textwidth}
   \includegraphics[width=\textwidth]{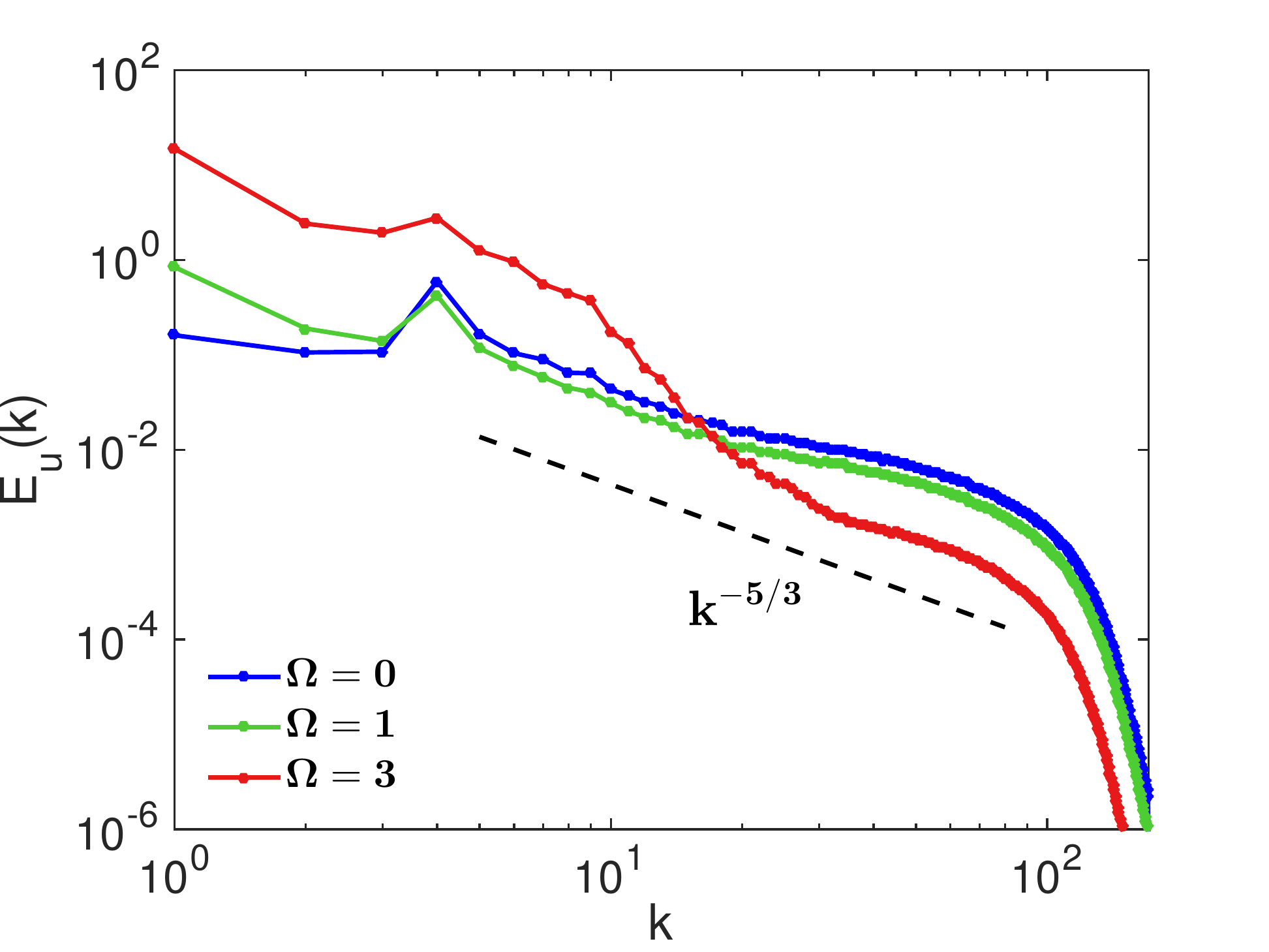}
   \caption{}
   \label{fig:Eu}
 \end{subfigure}
 \begin{subfigure}{0.5\textwidth}
   \includegraphics[width=\textwidth]{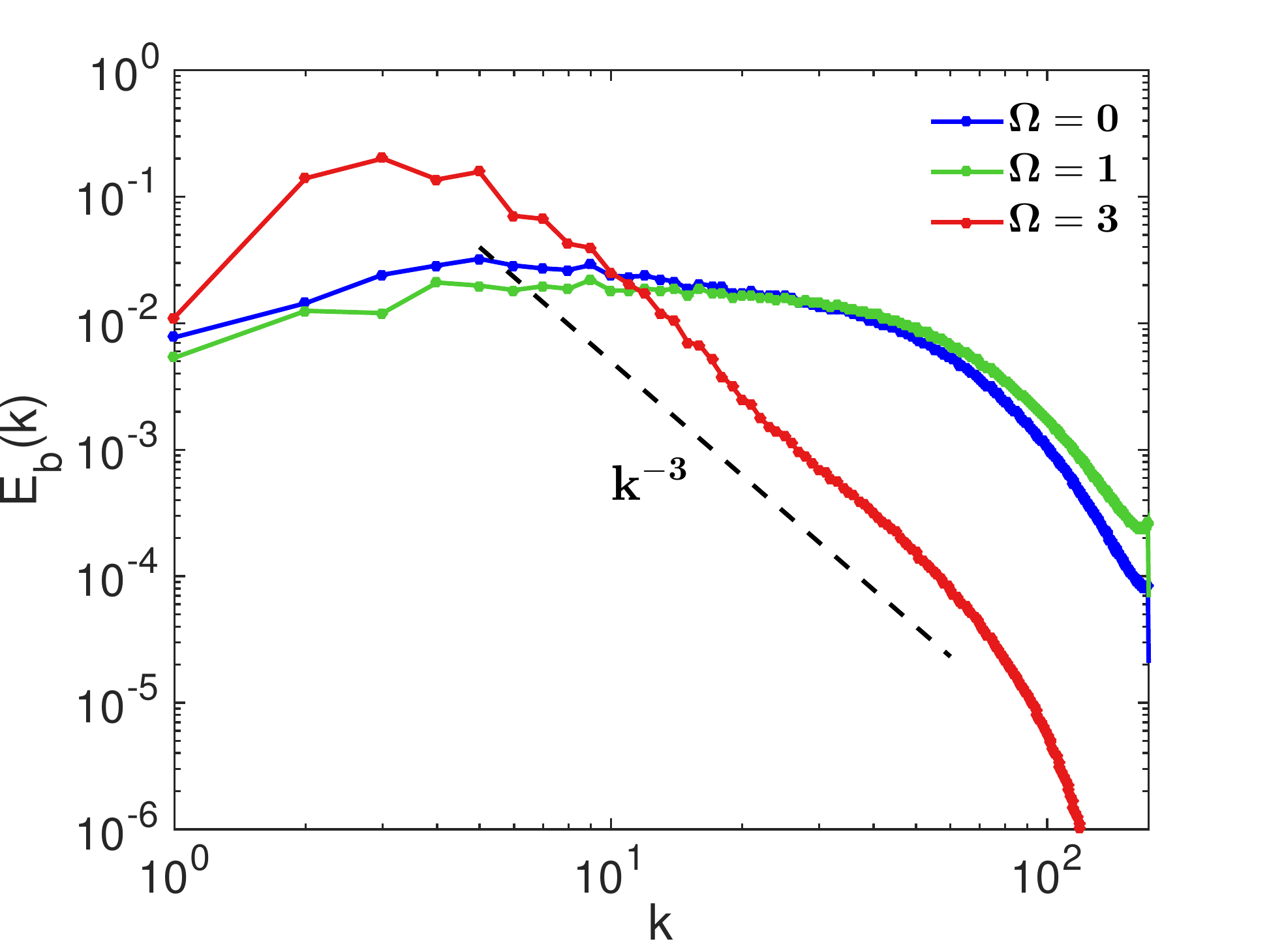}
   \caption{}
   \label{fig:Eb}
 \end{subfigure}
 \caption{(Color online) Spectra of a) the kinetic energy and b) the magnetic energy. Note that the $k^{-5/3}$ and the $k^{-3}$ curves are shown indicatively.} 
 \label{fig:spectra} 
 \end{figure}

In Fig. \ref{fig:Eu} we show the spectra of the kinetic energy for the flows with the three different rotation rates we considered (see Table \ref{tbl:parameters}). 
The flows with $\Omega = 0$ and $\Omega = 1$ fall in the regime I of Fig. \ref{fig:scalings} where behaviour close to Kolmogorov is observed with the spectrum $E_u(k) \propto k^{-5/3}$. 
Deviations from this scaling appear toward the dissipation range and are expected due to the use of hyperviscosity. The flow with $\Omega = 3$ falls in regime II of Fig. \ref{fig:scalings} where energy condensates at low $k$ changing the scaling of the spectrum. At large wavenumbers a spectrum close to $k^{-5/3}$ is expected to be recovered for length scales 
smaller than the Zeman scale, i.e. $k > (\Omega^3/\epsilon)^{1/2}$, which is the scale where the rotation period $\tau_w = \Omega^{-1}$ is equal to the eddy turnover time $\tau_{_{NL}}$ \citep{zeman94,hopfingeretal82}.

For reference, we also present the spectra of the magnetic energy $E_b(k)$ for $Rm$ close to onset (see Fig. \ref{fig:Eb}). For the flow cases $\Omega = 0$ and $\Omega = 1$ the spectra are almost flat with the magnetic energy equally distributed across a range of scales and with an exponential decay at high wavenumbers. On the other hand, the magnetic energy spectrum for the case  $\Omega = 3$ decreases strongly with $k$. The $k^{-3}$ power-law has been plotted as a guide to the eye. The peak of the magnetic energy is at $k = 3$, while the energy at the largest scale of the flow ($k = 1$) is more than an order of magnitude smaller.

\begin{figure}
 \centering
 \begin{subfigure}{0.49\textwidth}
   \includegraphics[width=\textwidth]{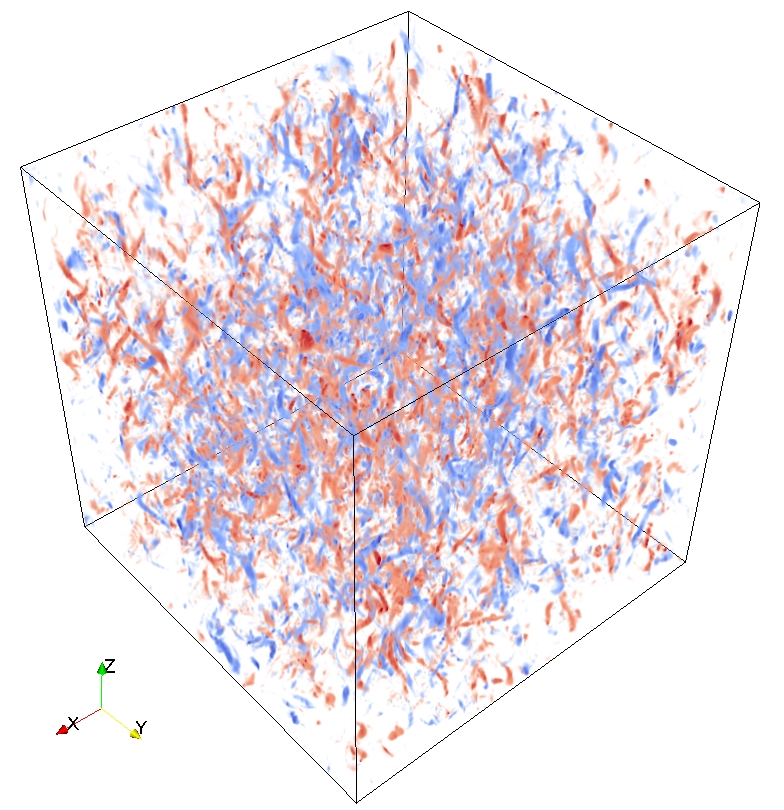}
   \caption{}
   \label{fig:omega0}
 \end{subfigure}
\begin{subfigure}{0.49\textwidth}
   \includegraphics[width=\textwidth]{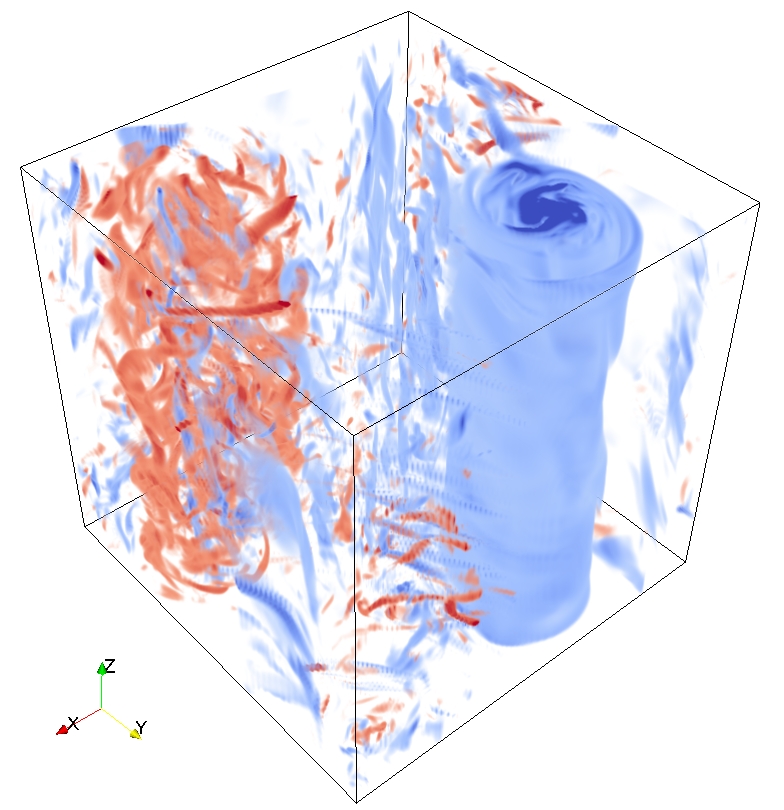}
   \caption{}
  \label{fig:omega3}
  \end{subfigure}
  \caption{(Color Online) Contour plots of the vertical vorticity $\omega_z$ for (a) $\Omega = 0$ (regime I) and (b) $\Omega = 3$ (regime II). Blue corresponds to positive values (co-rotating) and red to negative values (counter-rotating).}
  \label{fig:vizual}
 \end{figure}

In order to have a visual representation of the two flow regimes, we show renderings of the vertical component of vorticity $\omega_z$ in Fig. \ref{fig:vizual} for the flows with $\Omega = 0$ and $\Omega = 3$. The blue coloured contours correspond to vertical vorticity aligned with the background rotation ($\omega_z > 0$, co-rotating) while red correspond to vertical vorticity anti-aligned with the background rotation ($\omega_z < 0$, counter-rotating).
The non-rotating flow displays a vertical vorticity that is homogeneously distributed in the box over a large population of randomly oriented structures (see Fig. \ref{fig:omega0}). On the other hand, the rotating flow is organised into a strong coherent co-rotating large scale vortex and a counter-rotating vortex responsible for the energy cascade to small scales, which is typically observed in rotating turbulent flows as noted before \citep{dt16,alexakis15}. To sum up, from Figs. \ref{fig:spectra} and \ref{fig:vizual} we can infer two effects that take place: 
i) the suppression of turbulent fluctuations 
and ii) the organisation of the large scales in space and time, making the flow more efficient in driving the dynamo \citep{sda17}. 

\begin{figure}
 \begin{subfigure}{0.5\textwidth}
   \includegraphics[width=\textwidth]{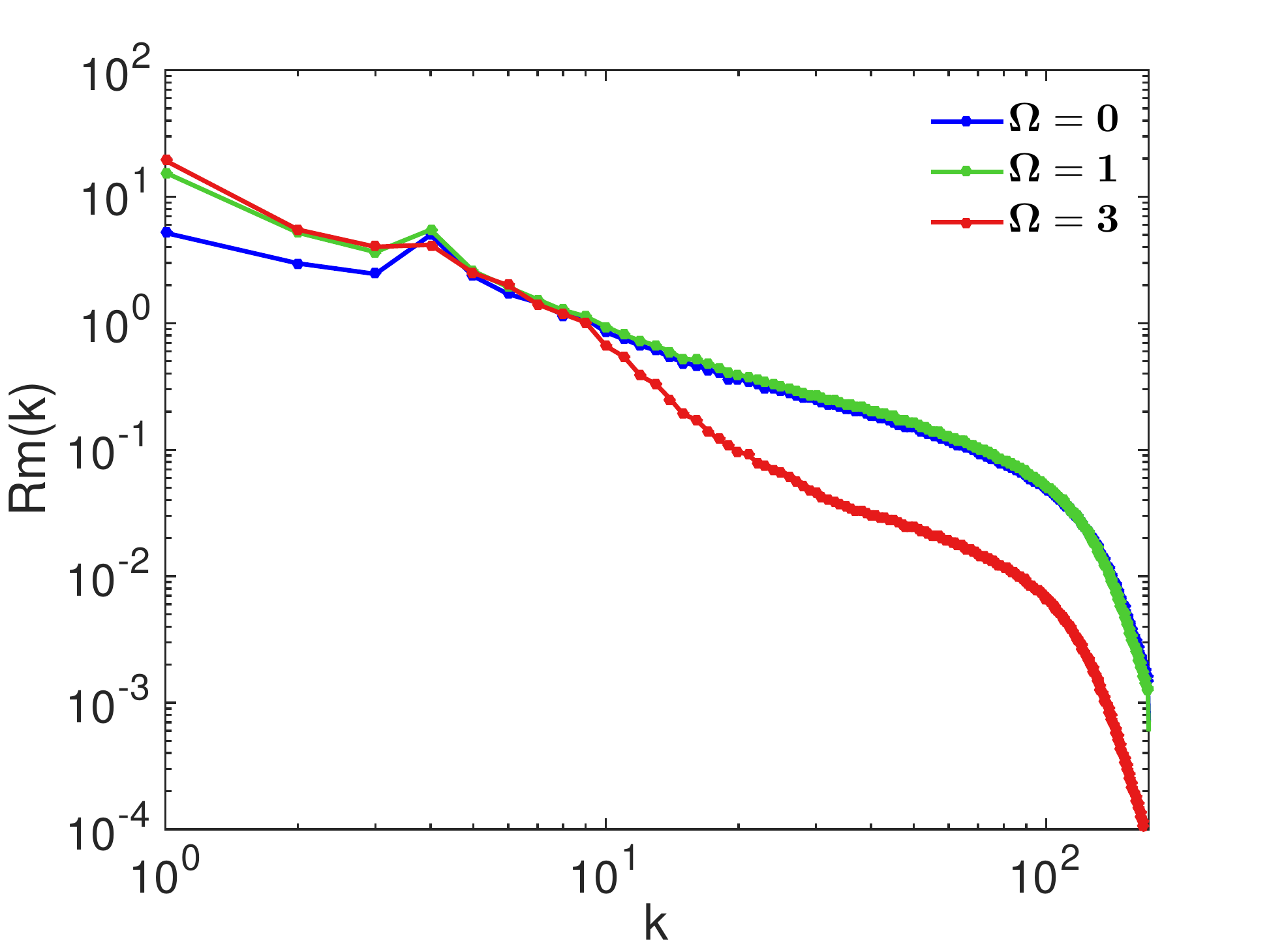}
   \caption{}
   \label{fig:Rm}
 \end{subfigure}
 \begin{subfigure}{0.5\textwidth}
   \includegraphics[width=\textwidth]{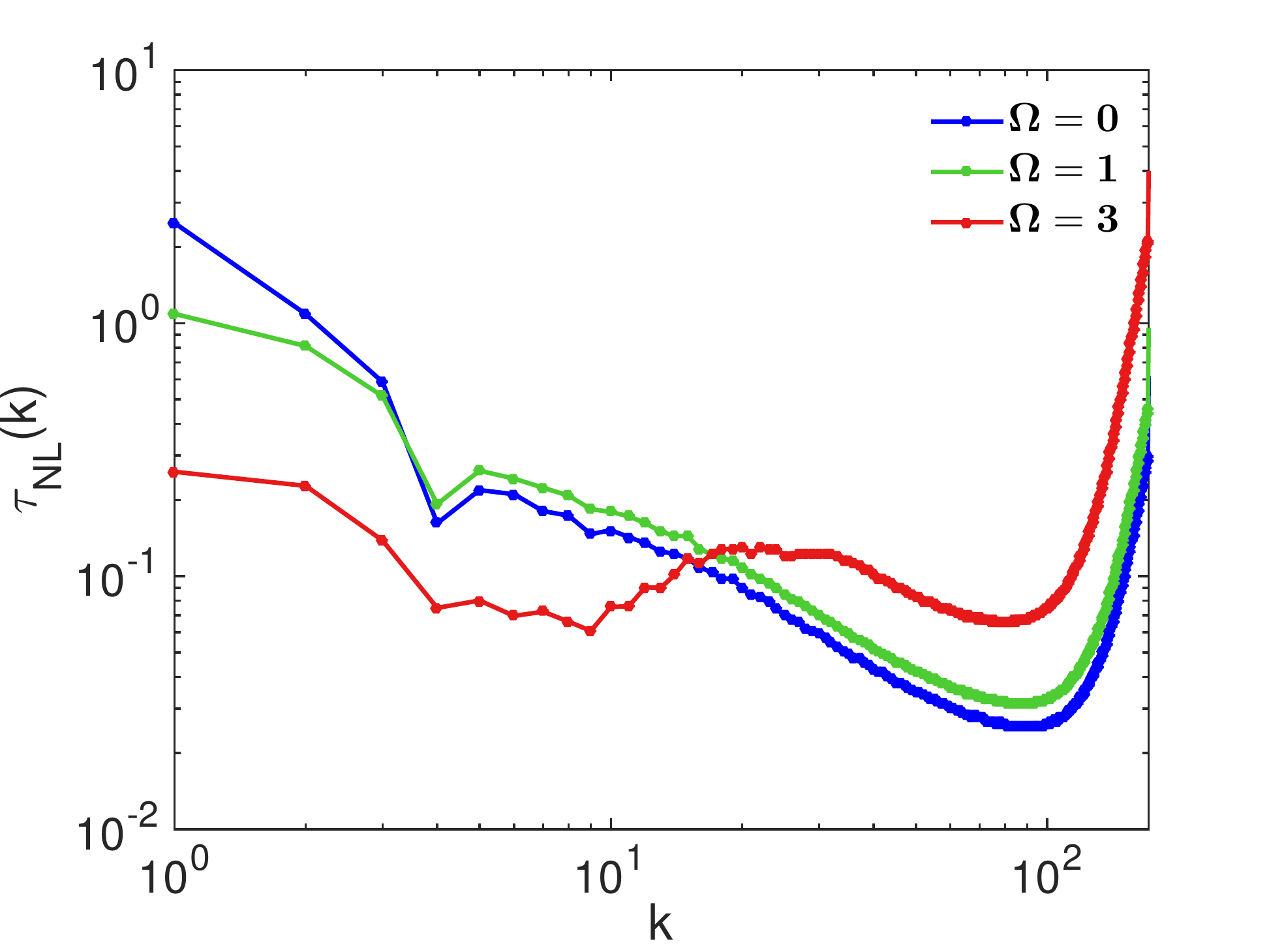}
   \caption{}
   \label{fig:taunl}
 \end{subfigure}
 \caption{(Color online) Wavenumber dependence of a) the magnetic Reynolds number and b) the turnover time scale}
 \label{fig:Rmtaunl}
 \end{figure}

To identify which of these effects and properties of the flow are more important for these low $Pm$ rotating dynamos we consider the wavenumber dependence of the magnetic Reynolds number (see Fig. \ref{fig:Rm}), which we compute based on the kinetic energy spectrum as
\begin{equation}
 Rm(k) = \frac{u(k)}{\eta_c k} = \lt(\frac{E_u(k)}{\eta_c^2 k}\rt)^{1/2},
 \label{eq:Rmk}
\end{equation}
using 
Eq. \eqref{eq:velk}.
For all the flows the magnetic Reynolds number decreases monotonically with $k$. Note that, for the cases with $\Omega = 1$ and $\Omega = 3$,  $Rm(k)$ is maximum at the largest scale of the flow ($k = 1$), while for $\Omega = 0$ is at the forcing scale ($k_f = 4$). Moreover, for the flow with $\Omega = 3$ the $Rm(k)$ spectrum is considerably suppressed for $k \geq 10$ in contrast to the other two flows. This observation indicates the suppression of the velocity fluctuations in this range of scales in agreement with Figs. \ref{fig:Eu} and \ref{fig:vizual}. 

The transfer of kinetic energy to magnetic energy occurs via the shearing of magnetic field lines and so the shear amplitude is a key quantity for dynamo action. Actually, the growth rate $\gamma$ is 
assumed to be proportional to the largest shear $S$ of the flow for a sufficiently complex flow with large enough $Rm$. Thus, ignoring any dissipative effects from dimensional arguments $S \propto U/\ell$ which is proportional to $\tau_{_{NL}}^{-1}$. So, let's now consider the scale-dependence of the eddy turnover time (see Fig. \ref{fig:taunl}), which we compute as
\begin{equation}
 \tau_{_{NL}}(k) = \frac{1}{k u(k)} = (k^3E_u(k))^{-1/2},
 \label{eq:taunlk}
\end{equation}
using Eq. \eqref{eq:velk}.
For the flows with $\Omega = 0$ and $\Omega = 1$ the smallest values of $\tau_{_{NL}}$ lie at the small scale eddies (see Fig. \ref{fig:taunl}) while for $\Omega = 3$ the smallest eddy turnover times occur at the wavenumber range $4 \leq k \leq 10$. In other words, for flows in regime II, which are flows with a coherent and a random component, the largest shear amplitudes occur at much larger scales than for random flows (i.e. flows in regime I), where the highest shear rates occur at small scales.

From the above observations is evident that the range of scales that determines the dynamo growth rate $\gamma$ depends on the slope of the kinetic energy spectrum because it controls the amplitudes of the local magnetic Reynolds number and the turnover time. For the flows in regime I is not clear whether the large scales which have the largest $Rm(k)$ but the longest $\tau_{_{NL}}(k)$ are more important than the small scales with the smallest $Rm(k)$ and the shortest $\tau_{_{NL}}(k)$. Even though for these random flows we cannot deduce which scales are going to determine the dynamo growth rate, for the flows in regime II the large scales are clearly those that have the first word on $\gamma$. This is because they exhibit the largest magnetic Reynolds number and the shortest turnover times (i.e. the largest shear amplitudes) across the scales.

\begin{figure}
\begin{subfigure}{0.5\textwidth}
  \includegraphics[width=\textwidth]{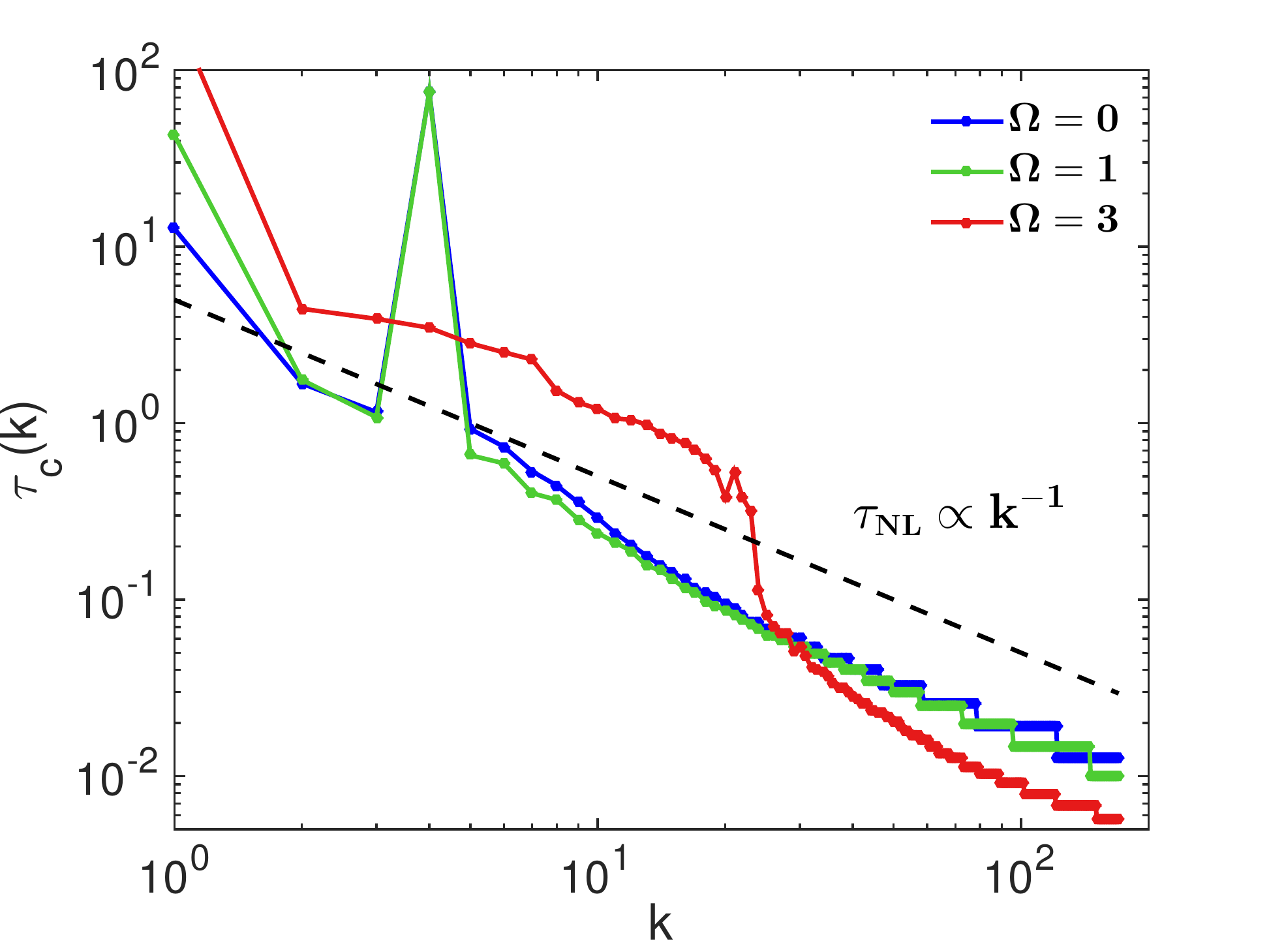}
  \caption{}
  \label{fig:tauc}
\end{subfigure}
\begin{subfigure}{0.5\textwidth}
   \includegraphics[width=\textwidth]{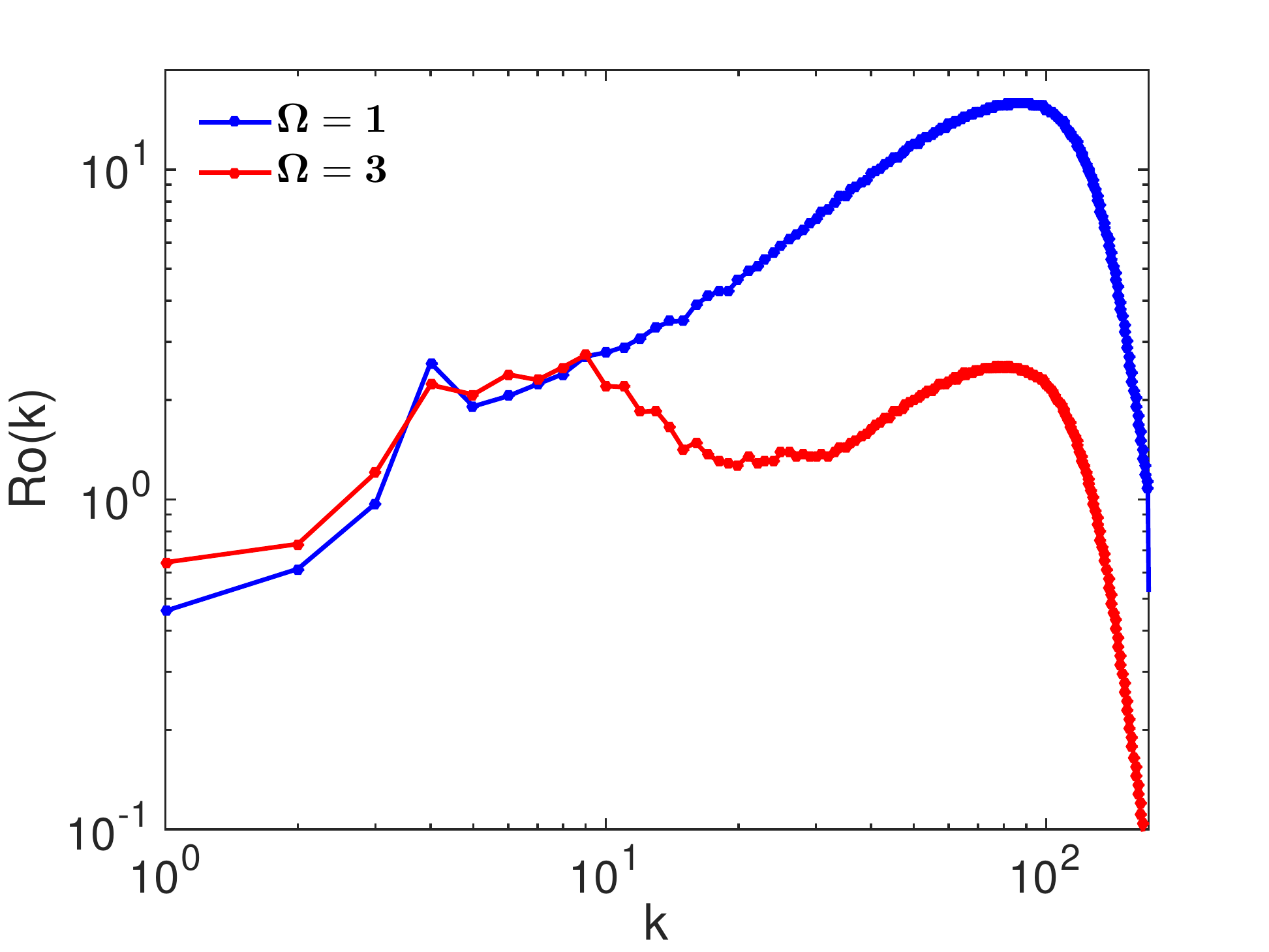}
  \caption{}
  \label{fig:Ro}
\end{subfigure}
\caption{(Color online) Wavenumber dependence of a) the correlation time scale and b) the Rossby number} 
\label{fig:taucRo} 
\end{figure}

Note though that besides the magnetic Reynolds number and the eddy turnover time, the dynamo growth rate is also a function of the coherence of the flow \citep{tobiascattaneo08b}. All of the above considerations are important only if the correlation time of the eddies is long compared with their turnover time. A measure that quantifies appropriately the coherence in the flow is the correlation timescale \citep{favieretal10}. To obtain the scale-dependence of the correlation timescale for our flows, we compute the Eulerian two-time correlation function of the velocity Fourier modes $\hat{\bf u}({\bf k}, t)$, which we define as
\begin{equation}
 R(k,\tau) = \frac{\avg{\hat{\bf u}({\bf k}, t)\hat{\bf u}^*({\bf k}, t + \tau)}}
 			 {\avg{\hat{\bf u}({\bf k}, t)\hat{\bf u}^*({\bf k}, t)}}
\end{equation}
where $^*$ indicates the complex conjucate modes of the velocity field and here the angle brackets $\avg{...}$ denote averages over the direction of the wave vector $\bf k$.
Then, from this correlation function 
we obtain the correlation time scale $\tau_c$ for each wavenumber $k$, which is defined as the half-width of the correlation function, i.e. $R(k,\tau_c) = 1/2$ for each $k$.
Figure \ref{fig:tauc} shows the scale dependence of the correlation time scale $\tau_c(k)$ for our flows with different rotation rates. The scaling of the 
nonlinear turnover time $\tau_{_{NL}} \propto (Uk)^{-1}$ has also been included in the plot for reference.
The correlation time of the eddies for the rotating flow with $\Omega = 3$ (regime II flow) is clearly larger for length scales up to the Zeman scale, i.e. $k \leq (\Omega^3/\epsilon)^{1/2}$ in comparison to the flows with $\Omega = 0$ and $\Omega = 1$ (regime I flows). On the other hand
the flow in regime II $\tau_c(k)$ is found to be smaller at lengthscales that do not feel the effect of rotation (i.e. $k > (\Omega^3/\epsilon)^{1/2}$) in contrast to the flows in regime I.

To identify those  scales that are most affected by the background rotation we plot the scale dependence of the Rossby number in Fig. \ref{fig:Ro}, which we have computed as
\begin{equation}
 Ro(k) = \frac{u(k) k}{2 \Omega} = \frac{(k^3E_u(k))^{1/2}}{2 \Omega}
\end{equation}
using Eq. \eqref{eq:velk}.
For weak rotation ($\Omega = 1$) the effect of rotation relative to the non-linear term becomes quickly negligible as scales become smaller. This is because $Ro(k)$ monotonically increases as $k$ increases until it reaches the highest wavenumbers where dissipation dominates. Now, for the flow with $\Omega = 3$ the effect of rotation across scales is much more important but still weakens when $Ro(k)$ monotonically increases again for large $k$. This is due to the suppression of fluctuations along the axis of rotation, which reduces the dissipation rate of the flow and thus the kinetic energy is distributed differently across scales as we saw in Fig. \ref{fig:Eu}. So, the plots in Fig. \ref{fig:taucRo} clealy suggest that the coherence in time of the flow is induced by the background rotation. It is this organised component of the flow whose coherence time is long compared with the turnover time and plays a decisive role on the vast improvement of the dynamo growth rate \citep{tobiascattaneo08a,sda17}.
To sum up,
using different measures we have identified that a range of large scales of the rapidly rotating turbulent flow, 
which exhibit i) large values of the magnetic Reynolds number, ii) small values of the eddy turnover time and iii) long coherence times, are those that determine the dynamo growth rate in comparison to random flows. 

 \section{The dynamo growth rate}

\begin{figure}
  \centerline{\includegraphics[width=0.5\textwidth]{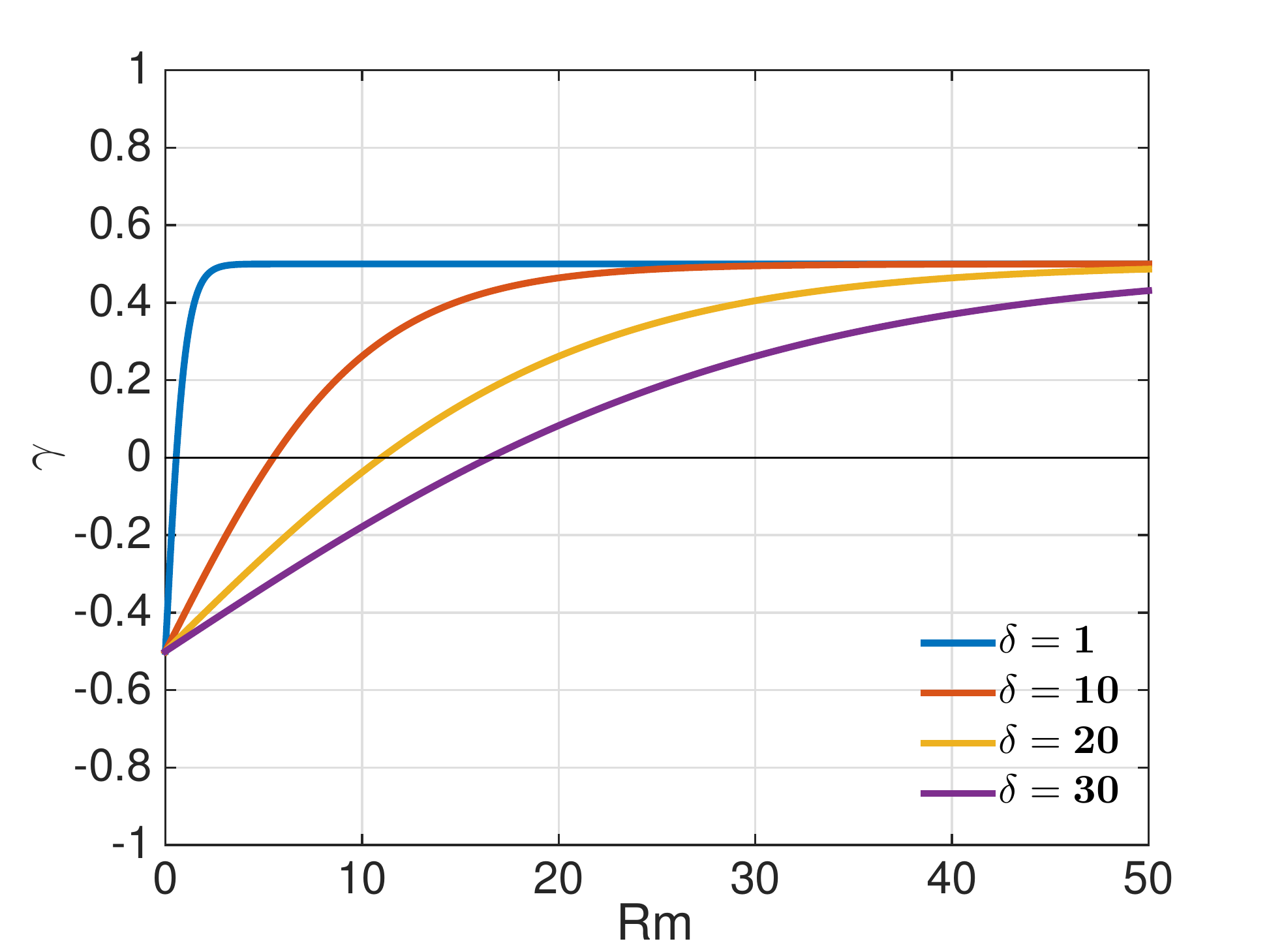}}
  \caption{(Color online) Model for the growth rate dependence on $Rm$}
  \label{fig:gmodel}
 \end{figure}

These ideas can be made somewhat more quantitative by assuming that each velocity scale acts in its own right as a `quick dynamo', i.e. that each dynamo scale reaches its maximum growth-rate quickly as a function of $Rm$ \citep{tobiascattaneo08a}. 
To be specific, we assume that 
the dynamo growth rate 
ascends very steeply  close to $Rm_c$. Following \cite{tobiascattaneo08a} we can model the growth rate $\gamma$ of such a dynamo at scale $k$, where $\gamma$ is measured in terms of the (inverse) of the local turnover time, by assuming the following growth-curve dependence on $Rm$
\begin{equation}
 \gamma = \gamma_{min} + (\gamma_{max} - \gamma_{min}) \tanh(Rm/\delta)
 \label{eq:gmodel}
\end{equation}
where $\gamma_{min}$ is the (negative) growth rate at $Rm = 0$, $\gamma_{max}$ is the maximum growth rate and $\delta$ is a fitting parameter that gives the dependence of growth rate on $Rm$. In figure \ref{fig:gmodel} we plot Eq. \eqref{eq:gmodel} for different values of $\delta$ and for $\gamma_{min} = -0.5$ and $\gamma_{max} = 0.5$.

\begin{figure}
 \begin{subfigure}{0.32\textwidth}
   \includegraphics[width=\textwidth]{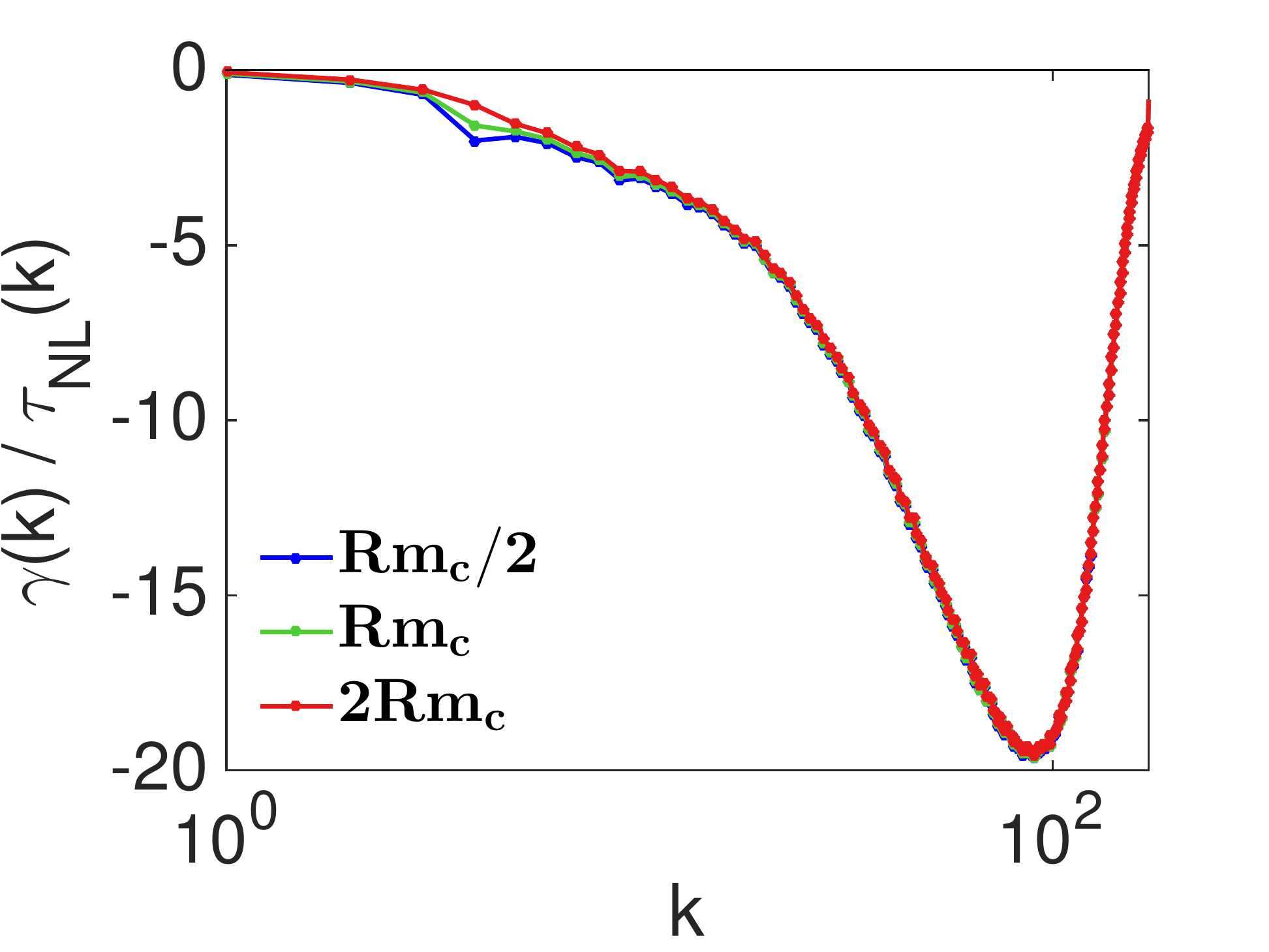}
   \caption{}
   \label{fig:gamma_a}
 \end{subfigure}
 \begin{subfigure}{0.32\textwidth}
   \includegraphics[width=\textwidth]{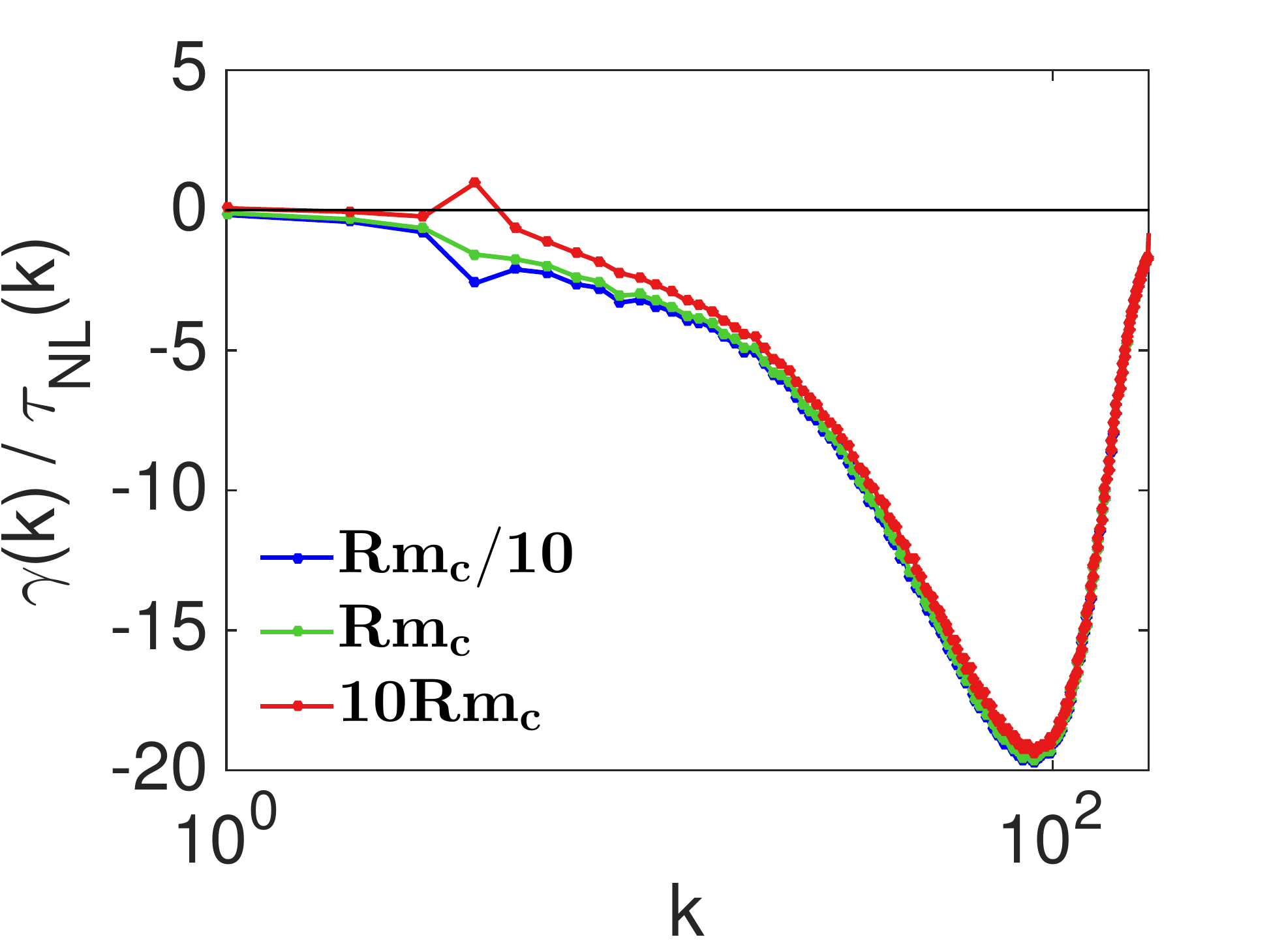}
   \caption{}
   \label{fig:gamma_b}
 \end{subfigure}
\begin{subfigure}{0.32\textwidth}
   \includegraphics[width=\textwidth]{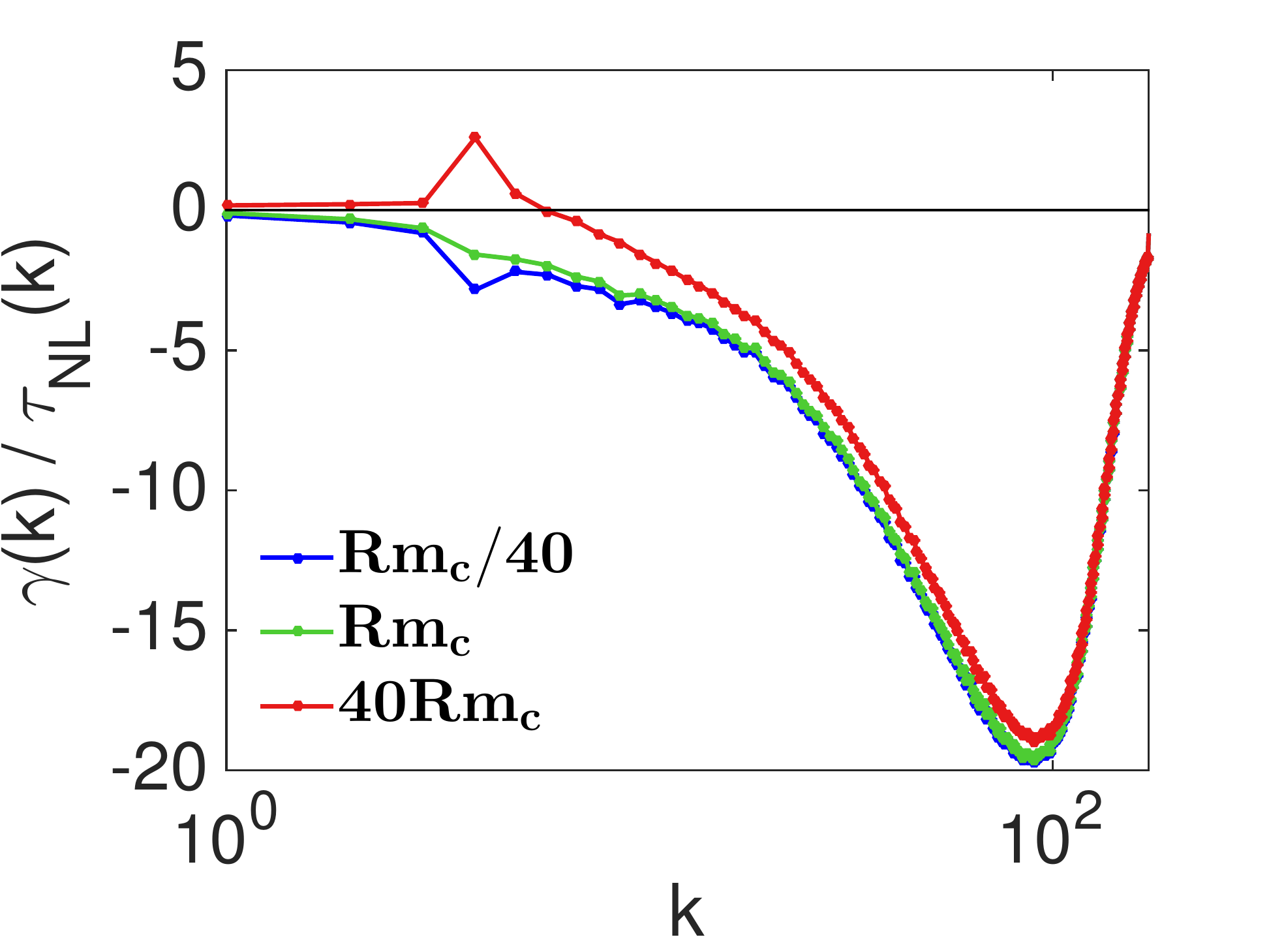}
   \caption{}
   \label{fig:gamma_c}
 \end{subfigure} \\
 \begin{subfigure}{0.32\textwidth}
   \includegraphics[width=\textwidth]{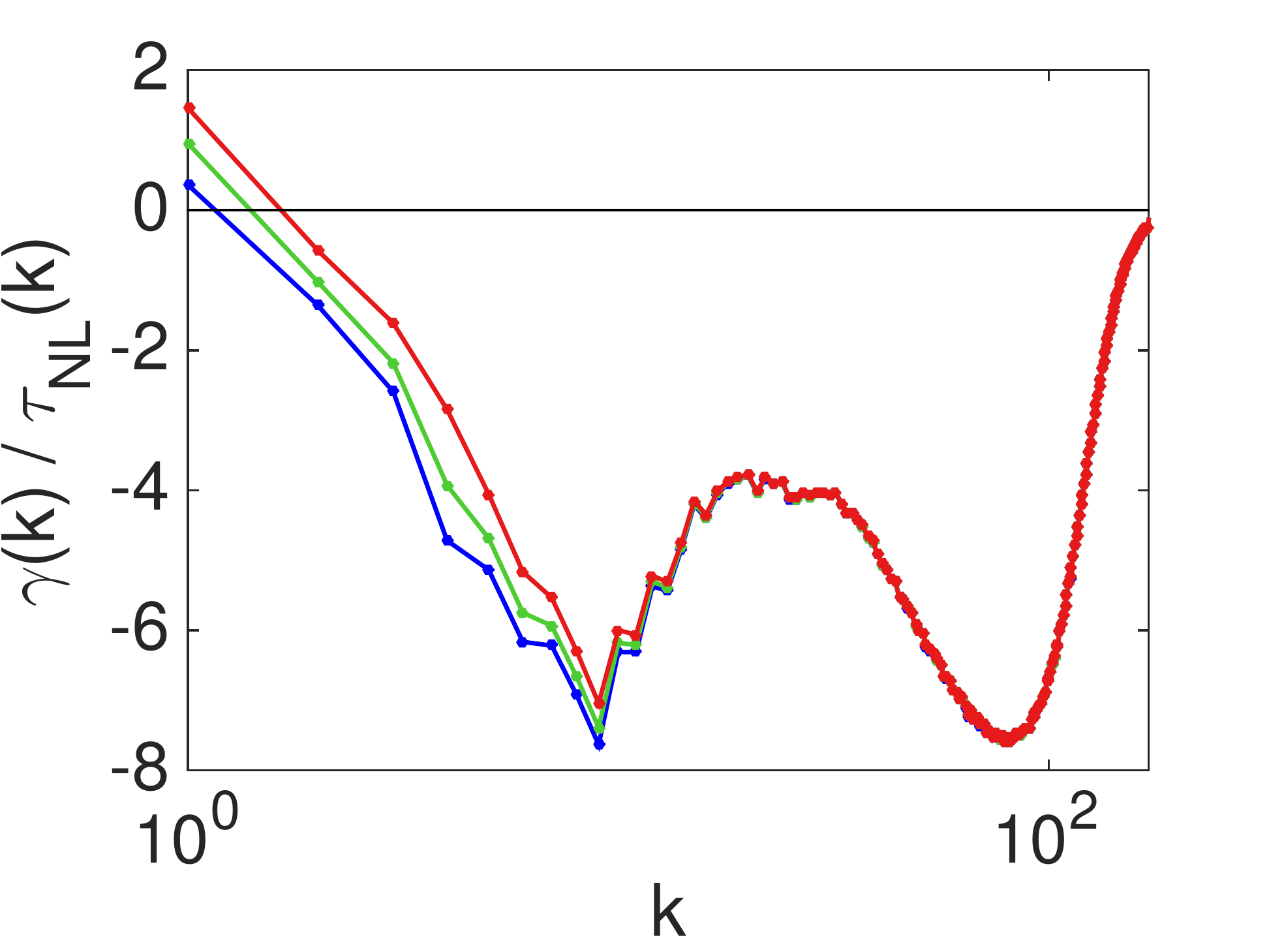}
   \caption{}
   \label{fig:gamma_d}
 \end{subfigure}
 \begin{subfigure}{0.32\textwidth}
   \includegraphics[width=\textwidth]{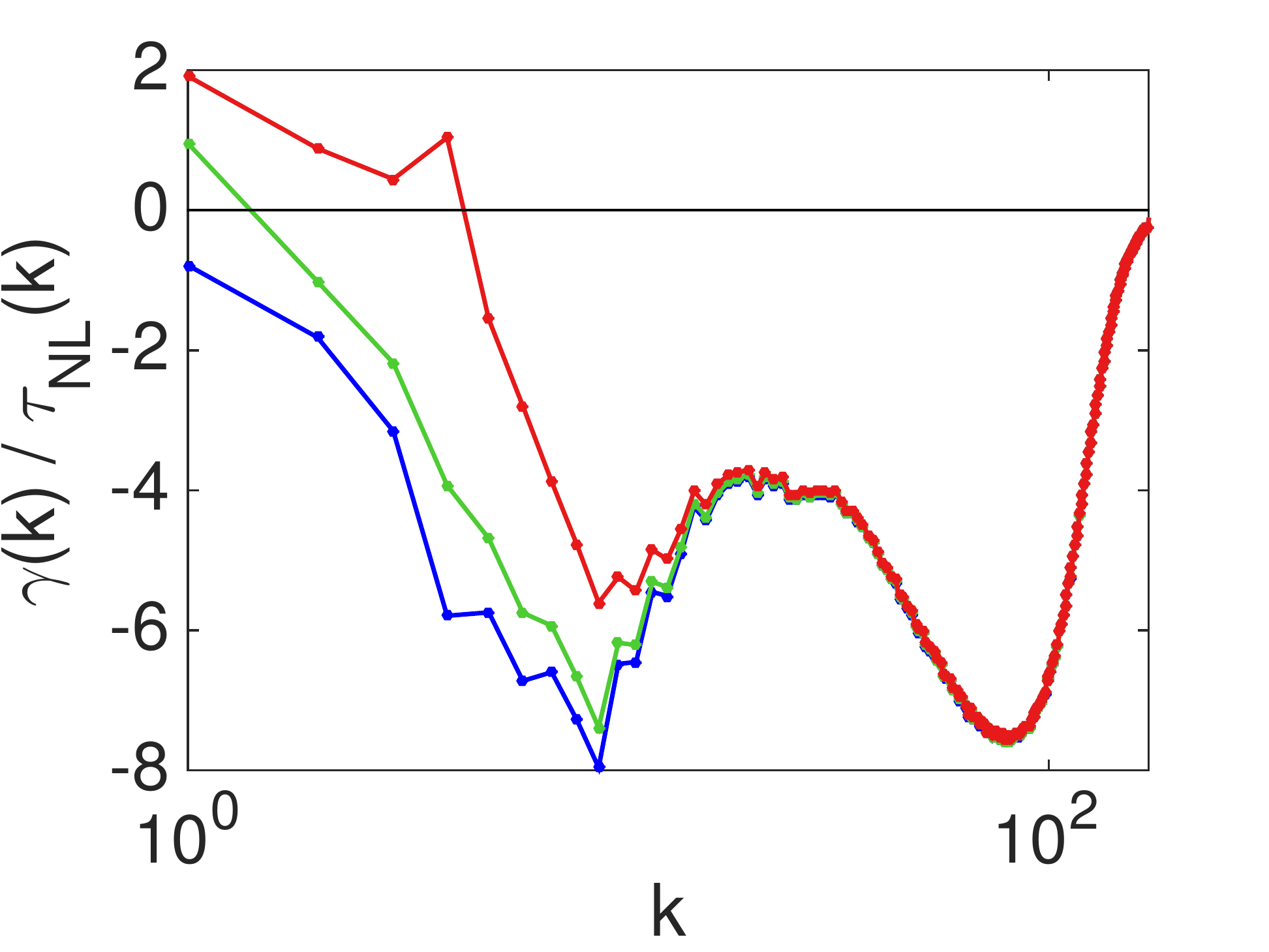}
   \caption{}
   \label{fig:gamma_e}
 \end{subfigure}
\begin{subfigure}{0.32\textwidth}
   \includegraphics[width=\textwidth]{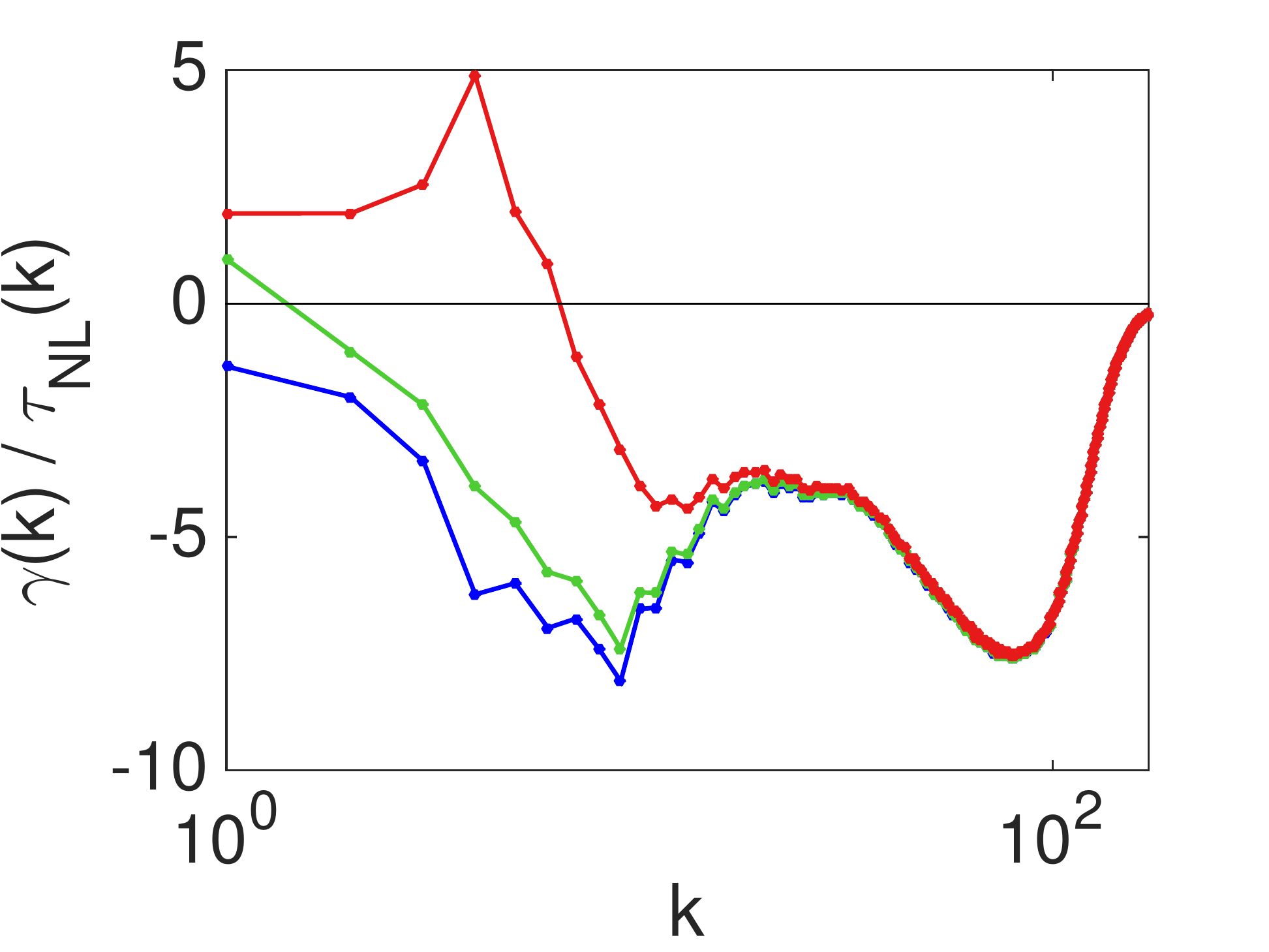}
   \caption{}
   \label{fig:gamma_f}
 \end{subfigure}
  \caption{(Color online) Scale-dependence of the growth rate $\gamma$ per eddy turnover time $\tau_{_{NL}}(k)$. Top row: $\Omega=0$ (regime I). Bottom row: $\Omega=3$ (regime II).}
  \label{fig:growthrate}
 \end{figure}

In this scenario each scale in isolation can act as a dynamo, with dynamo action setting in at $Rm \sim \mathcal O(1)$ and reaching its maximum growth rate by $Rm \sim 30-50$. This is not unreasonable for fast dynamo action \citep{gallowayproctor92,seshasayanan2016b}. So, without loss of generality we choose $\delta = 20$, $\gamma_{min} = -0.5$, $\gamma_{max} = 0.5$ and we plot the scale dependence of the growth rate $\gamma(k)$ using Eqs. \eqref{eq:gmodel} and \eqref{eq:Rmk} divided by the local turnover time from Eq. \eqref{eq:taunlk} to identify the dynamo scales (see Fig. \ref{fig:growthrate}). 
The top row of Fig. \ref{fig:growthrate} corresponds to the flow with $\Omega = 0$, while the bottom row to $\Omega = 3$. In this figure we also plot the scale dependence of the growth rate per local turnover time using subcritical ($Rm < Rm_c$) and supercritical ($Rm > Rm_c$) values of the magnetic Reynolds number in Eq. \eqref{eq:gmodel}. In this way, we want to show the scale dependence of the growth rate on $Rm$ by keeping $Pm$ fixed. Any points of the curves lying above the $\gamma(k)/\tau_{_{NL}}(k) = 0$ line indicate the dynamo scales.

If we now compare Figs. \ref{fig:gamma_a} and \ref{fig:gamma_d} we see that there are no dynamo scales for the flow with $\Omega = 0$, while $k=1$ turns out to be a dynamo scale for the flow with $\Omega = 3$. 
From Fig. \ref{fig:growthrate} is clear that by changing the value of $Rm$ by an order of magnitude in either direction when we compute $\gamma(k)$ does not affect the dynamo scales much. This is 
particularly true for the non-rotating flow (see top row in Fig. \ref{fig:growthrate}) where dynamo scales appear only for values of $Rm$ much greater than $Rm_c$. On the contrary, for the flow with $\Omega = 3$ it is clear that as $Rm$ increases further from $Rm_c$ the number of wavenumbers with $\gamma(k)/\tau_{_{NL}}(k) > 0$ increases cosiderably and the scale with the largest growth rate becomes smaller (see Figs. \ref{fig:gamma_e} and \ref{fig:gamma_f}). Finally, the general picture from Fig. \ref{fig:growthrate} is that the flow in regime II is a better dynamo than the flow in regime I, in the sense that a lot more scales have positive and larger growth rates per local turnover for the rotating flow than for the non-rotating flow.

\section{The importance of  coherence for the dynamo growth rate}

In this section we shall test the assertion of the previous section that it is the modification of the coherence time of the turbulence in the rotating system that is responsible for making the dynamo at low $Pm$ more efficient. In order to do this we construct a numerical algorithm designed to  compare dynamos with the same spectra but different coherence time.

We consider the hydrodynamic flow  with $\Omega = 3$ (recall that for the kinematic regime the flow evolves independently of the state of the magnetic field). This flow has a well-established spectrum and therefore nonlinear time. Before inputting this flow into the induction equation,  we randomise the phases of each Fourier coefficient (keeping the amplitude fixed) on a timescale that we identify as the coherence time scale $\tau_c$. We adopt  the randomisation procedure given by ${\bf \hat u}_{new}({\bf k_\perp}) = {\bf \hat u}({\bf k_\perp})\exp(i\phi_{k_\perp})$, where $\phi_{k_\perp}$ are random numbers that depend only on the horizontal wavenumber $k_\perp = \sqrt{k_x^2 + k_y^2}$. This procedure essentially shifts the flow pattern (including the  large scale vortex) to a random position on a given timescale $\tau_c$. We choose five different (normalised)  coherence times  $\tau_c/\Delta t$ as Fig. \ref{fig:gammatauc} illustrates. Note that the square symbol denotes the flow without randomised phases, i.e. where $\tau_c$ is given by the long-lived coherence of the rotationally constrained flow. Moreover the flow with $\tau_c = \Delta t$ denotes a delta-correlation in time. Figure \ref{fig:gammatauc} shows the growth-rate of the kinematic dynamo as the correlation time is decreased.
\red{ \begin{figure}
  \centerline{\includegraphics[width=0.5\textwidth]{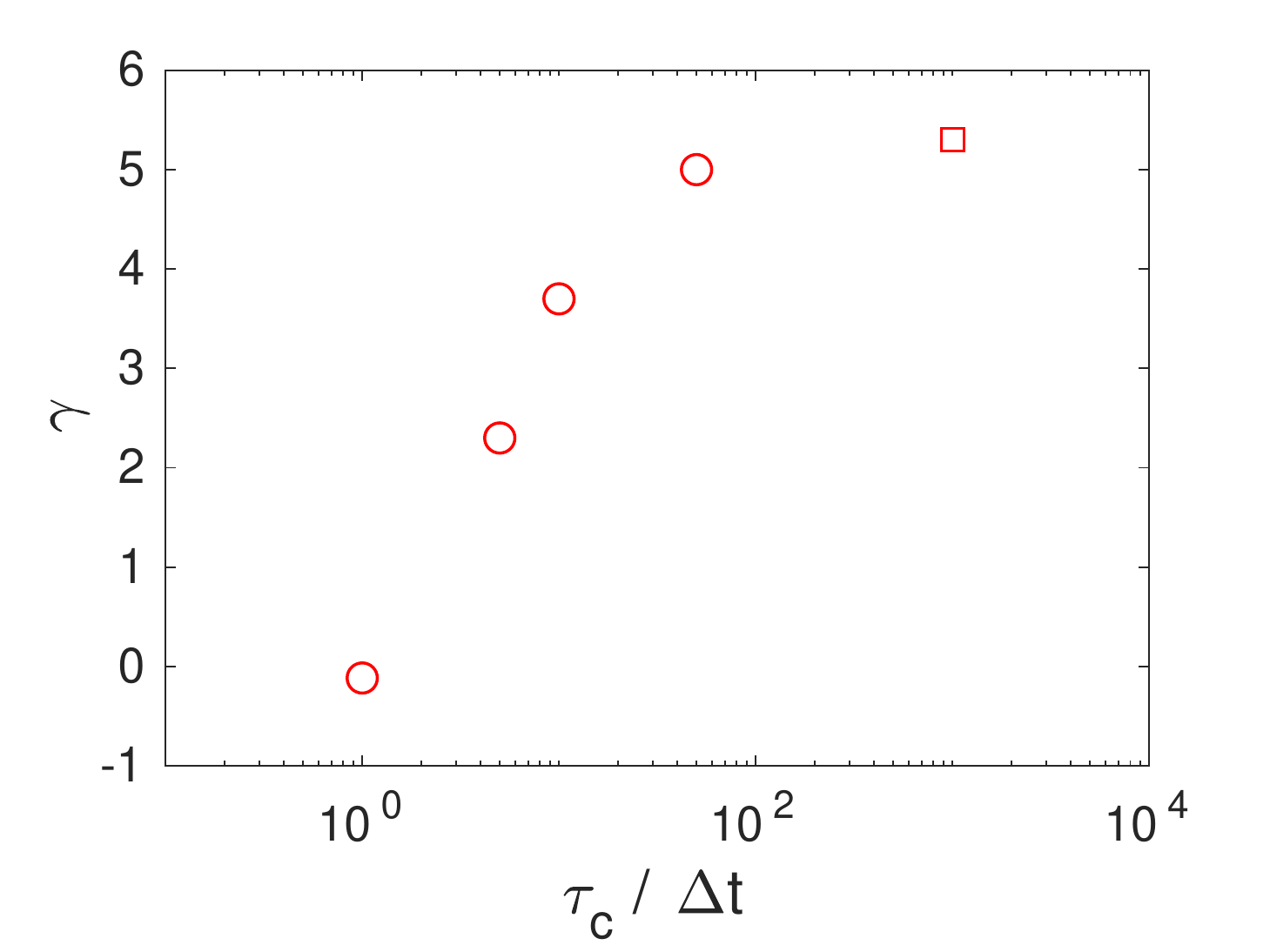}}
  \caption{(Color online) Dependence of the dynamo growth rate on the correlation time scale $\tau_c$ for the rotating flow with $\Omega=3$ (regime II). Note that the square symbol denotes the flow without randomised phases, i.e. $\tau_c = \infty$.}
  \label{fig:gammatauc}
 \end{figure}}

It is clear that as the ratio $\tau_c/\Delta t$ decreases, and thus the large scale vortex becomes less coherent in time, the growth rate $\gamma$ decreases. Recall that the slope of the spectrum of the flow is the same for all cases at all scales (most importantly at the dissipative scale of the magnetic field). We note that, for this case, $dt \simeq 0.15$ and so the growth-rate deviates significantly from the undisturbed (long correlation time) case when the decorrelation time $\tau_c \simeq 4.5$. Comparing this timescale with those obtained from Fig. \ref{fig:tauc}, it is clear that the growth-rate starts to drop when the imposed decorrelation timescale becomes smaller than that on which the large-scale vortex naturally decorrelates.
The figure confirms the conjecture that it is the increase in the correlation time that increases the effectiveness of the rotating dynamo and reduces the critical Rm for dynamo onset.

 \section{Conclusions}

One of the fundamental questions in kinematic dynamo theory is what determines the growth rate $\gamma$ of the dynamo instability, and how. For random flows like the Kazantsev-Kraichnan model, at low $Pm$, it has been shown that $\gamma$ is determined by the slope of the energy spectrum and that the growth time is of the order of the turnover time at the scales with the highest shear amplitude, which are the resistive scales in such flows. However, geophysical and astrophysical flows contain a coherent component and whether the theory from the Kazantsev-Kraichnan model is enough to explain what determines the dynamo growth rate in these flows is an open question. 

In this paper, we have addressed this question by analysing flows in two regimes. 
Regime I consists of the non-rotating and slowly rotating flows, which can be considered to be in some sense random and regime II consists of the rapidly rotating flows, which contain a significant coherent component reminiscent to the structures of geophysical and astrophysical flows. For rapidly rotating flows we observe two effects:
i) the suppression of turbulent fluctuations along the axis of rotation and 
ii) the organisation of the large scales in space and time. 
The impact of these two effects on flows in regime II is the change of the kinetic energy spectrum which has a clear influence on the local magnetic Reynolds number $Rm$ and turnover time $\tau_{_{NL}}$. Analysing the scale dependence of these two quantities, we find that the large scales of the flow in regime II exhibit large values of $Rm$ and small values of $\tau_{_{NL}}$, which benefit the dynamo onset. 

At this point, one can claim that the Kazantsev-Kraichnan model could be enough to explain the growth rate because all we get is essentially a change of the slope of the kinetic energy spectrum. However, by computing the scale dependence of the correlation timescale we also find that the large scales are dominated by coherence times much longer than the local turnover times.
Finally we have performed a numerical experiment for our rapidly rotating flow where the flow is decorrelated on a timescale $\tau_c$ before being input into the induction equation. This experiment demonstrates clearly that the primary effect of rotation in modifying the dynamo is via an increase in the correlation time of the flow.
Hence, the Kazantsev-Kraichnan formalism is not useful and one has to appeal to the multi-scale quick-dynamo theory of \citet{tobiascattaneo08a}. Thus, the combination of i) the long coherent times and ii) the large values of $Rm$ 
at the large scales are the characteristics that determine the dynamo growth rate in the rapidly rotating turbulent flow. 
%
Our analysis should be applicable to more general geophysical and astrophysical flows that are influenced by rotation, stratification and shear as these flows tend to be dominated by coherent structures. This is something that should be tested further from future studies.

\begin{acknowledgements}
V.D. acknowledges support from the Royal Society and the British Academy of Sciences (Newton International Fellowship, NF140631). The computations were performed using ARC, the High Performance Computing facilities at the University of Leeds, UK. We would also like to thank an anonymous referee for his/her comments that made us improve our manuscript.
\end{acknowledgements}

\bibliographystyle{jfm}
\bibliography{references}
\end{document}